\documentclass[prd,showpacs,preprintnumbers,twocolumn,superscriptaddress,floatfix]{revtex4-2}
\usepackage[utf8]{inputenc}
\usepackage{bm,amsmath,amssymb}
\usepackage{graphicx}
\usepackage{subfigure}
\usepackage{color}
\usepackage{hyperref}
\usepackage{multirow}
\usepackage{soul}
\usepackage{amssymb}

\let\oldalign\align
\def\align{\linenomath\oldalign}

\begin{document}

\title{Revisiting the sphaleron and axion production rates in QCD at high temperatures }

\author{Sayak Guin}
\email{sayakg@imsc.res.in}
\affiliation{The Institute of Mathematical Sciences, Chennai 600113, India}
\affiliation{Homi Bhabha National Institute, Training School Complex, Anushaktinagar, Mumbai 400094, India}

\author{Sayantan Sharma}
\affiliation{The Institute of Mathematical Sciences, Chennai 600113, India}
\affiliation{Homi Bhabha National Institute, Training School Complex, Anushaktinagar, Mumbai 400094, India}

\begin{abstract}

We report our new lattice results for the sphaleron rate calculated within a thermal effective field theory of soft 
SU(N) gluons whose momenta are below the magnetic scale, where $N=2,3$, for a wide range of temperatures spanning from 
$0.6$ to $10^{15}$ GeV at sufficiently large volumes. Comparing these results with sphaleron rates in a nonthermal SU(N) 
plasma where the infrared gluons are overoccupied, we estimate the typical thermalization time for these ultrasoft gluons during the early stages of reheating after inflation. We also calculate the thermal production rate of 
relativistic axions due to these non-perturbatively interacting soft gluons which shows a significant deviation from its 
perturbative estimate even at the electroweak scale. 
    
\end{abstract}

\maketitle


\section{Motivation \& Outline}
In strongly interacting nuclear matter described by Quantum Chromodynamics (QCD) the axial current 
$j^{\mu}_{5,f}$ of the quarks is not conserved due to quantum fluctuations. For each flavor $f$ 
of quarks in the fundamental representation of the gauge group with $N$ colors, this violation 
is described by the anomaly relation,
\begin{equation}
    \partial_{\mu}j^{\mu}_{5,f}=2m\Bar{q}_{f}i\gamma_5q_f-\frac{g^2}{16\pi^2}G^c_{\mu\nu}\tilde{G}^c_{\mu\nu}~.
    \label{eqn:anomalyRel}
\end{equation}
where $G^c_{\mu\nu}$ represents the field strength tensor, 
$\tilde{G}^c_{\mu\nu}=\frac{1}{2}\varepsilon_{\mu\nu\alpha\beta}G^c_{\alpha\beta}$ is its dual, 
$q_f$ are the quark fields and $g$ is the gauge coupling. The color index runs from $c=1$-$N^2-1$. 
For massless quarks i.e. $m=0$ the right hand side of Eq.~\ref{eqn:anomalyRel} 
can be written in terms of the divergence of the Chern-Simons current, $K_\mu$, where
\begin{equation}
     K^{\mu}(x)=\frac{g^2}{32\pi^2}\epsilon^{\mu\nu\rho\sigma}\left(A_{\nu}^cF^c_{\rho\sigma}(x)-\frac{g}{3}f_{cde}
A_\nu^cA_\rho^dA_\sigma^e(x)\right)~.
\end{equation}
such that the Chern-Simons number $N_\text{CS}$ is the corresponding charge  
\begin{equation}
    N_\text{CS}(t)=\int d^3\mathbf{x}~K^0(t,\mathbf{x})~.
\end{equation}
A unique property of this $N_\text{CS}$ for non-Abelian gauge theories is that it is an integer  
and labels each degenerate vacuum state. Hence, each vacuum state of QCD is distinct and 
topologically inequivalent. The tunneling solutions between two vacua are known 
as instantons~\cite{Belavin:1975fg,tHooft:1976snw} which are characterized by an integer 
topological charge corresponding to the difference in the $N_\text{CS}$ labeling these vacua.  
When the typical energy fluctuations are larger in magnitude than the height 
$\sim \Lambda_{\text{QCD}}$ of the barrier that separates two topologically distinct vacua, 
certain gauge field solutions will exist which represent a roll-over from the top of the barrier 
to a vacuum state. Such a solution is known as sphaleron~\cite{Klinkhamer:1984di}. The rate of 
change of $N_\text{CS}$ with time due to topological transitions can be denoted as,
\begin{equation}
    \frac{dN_\text{CS}}{dt}=\frac{g^2}{8\pi^2}\sum_{c=1}^{N^2-1}\int d^3\mathbf{x}~ E_{i}^c(\mathbf{x})B_i^c(\mathbf{x})~.
    \label{eqn:CSnumberChange}
\end{equation}
The autocorrelation of the Chern-Simons number at two different times $t$ and $t+\Delta t$ 
given by
\begin{equation}
    \Gamma_{\text{sph}}=\lim_{V\to \infty,  t\rightarrow \infty}\frac{\Big\langle\left[N_\text{CS}(t+\Delta t)-N_\text{CS}(t)\right]^2\Big\rangle}{V\Delta t}
\end{equation}
thus denotes the sphaleron transition rate $\Gamma_{\text{sph}}$ assuming that change in the 
Chern-Simons number is diffusive at long enough time evolution $t\to \infty$. Accurately determining 
the sphaleron rate in QCD is important for many important physical phenomena. Sphaleron rates 
contribute to the damping of the coherent oscillations of axions~\cite{McLerran:1990de}, to exotic transport 
phenomena, e.g. the chiral magnetic effect~\cite{Kharzeev:2024zzm} and are a source of strong CP violation which is believed
to have played a role during baryogenesis in the early
Universe~\cite{Mohapatra:1991bz}.

It was anticipated early-on that the sphaleron rate can be measured controllably in classical SU(N) 
theory~\cite{Grigoriev:1988bd} at finite temperatures $T$. However it was soon realized that 
classical gauge theories in 3+1 dimensions suffer from ultraviolet divergences and the hard gluons 
with momenta $|\mathbf{p}| \gtrsim \pi T$ play an important role in the estimation of $\Gamma_{\text{sph}}$. 
These hard gluons, though not included in the classical theory, influence the sphaleron dynamics since the 
hard scale sets the momentum cut-off and thus the relaxation rates in the effective theory~\cite{Arnold:1999uy}. 
Incorporating this fact properly leads to a $ \Gamma_{\text{sph}} \propto g^{10}T^4$ rather than the 
naive expectation $\sim g^8T^4$ that comes just from dimensional analysis. A more careful analysis of the 
sphaleron rate leads to a dependence of the form $\sim \log(1/g)g^{10} T^4$~\cite{Bodeker:1998hm} arising 
due to the logarithmic dependence of the Debye mass on $g$. In the limit when $\log(1/g) \gg 1$, an effective 
field theory of the soft modes of the non-Abelian gauge theory at leading logarithmic order can be 
constructed~\cite{Bodeker:1998hm} which remains valid even at next-to-leading-log order 
if the color conductivity which goes as an input parameter is also calculated at the same order~\cite{Arnold:1999uy}. 
Furthermore, this effective theory is not plagued by ultraviolet divergences which is inherent in the classical 
Hamiltonian for non-Abelian gauge theory~\cite{Bodeker:1995pp}. The $\Gamma_{\text{sph}}$ measured on the 
lattice~\cite{Moore:2010jd} within this effective theory of soft gluons provides the leading $\log(1/g)$ dependence 
at sufficiently weak couplings, $g\lesssim 0.6$. 

Sphaleron transitions are also ubiquitous in a gluonic plasma under far-from-thermal equilibrium 
conditions, where the relevant scale $Q_s$, which denotes the gluon saturation scale, is larger than the height 
of the barrier between two vacua i.e. $Q_s \gg \Lambda_{\text{QCD}}$. A typical realization of a non-thermal 
state in SU(N) gauge theory consists of infrared gluons whose occupation numbers in a particular momentum state 
are large i.e. $1/\alpha_s(Q^2)\gg 1$, which allow for a classical-statistical description of such 
modes~\cite{Mace:2016svc}. 
Starting from such an initial state, the classical Hamiltonian evolution leads to a self-similar scaling regime, 
where the momentum distribution function of gluons reaches a stationary steady 
state~\cite{Schlichting:2012es,Berges:2013eia,berges2014universal}. 
Magnetic, electric and hard scales can be defined in such a system as well, and a clear scale hierarchy exists 
in this self-similar regime analogous to a thermal non-Abelian plasma at high enough temperatures~\cite{Berges:2023sbs}. 
Sphaleron transitions in such a self-similar non-thermal regime can be described as 
a random walk of $N_\text{CS}$ values with time, whose rate has a $(Q_s t)^{-
\frac{4}{3}}Q_s^4$ dependence on the gluon saturation scale $Q_s$ for SU(2) gauge theory~\cite{Mace:2016svc}. 
This rate is significantly higher than in a thermal plasma of gluons with a similar energy density. 

In this work we re-visit the calculation of the $\Gamma_{\text{sph}}$ in SU(N) gauge theories using lattice 
techniques both in-and-out-of-thermal equilibrium conditions where the $N=3$ results for the non-thermal case 
are new and not discussed earlier in the literature. We work in an effective theory of the soft gluons whose 
momenta are of the order of the magnetic scale and lower~\cite{Bodeker:1998hm} which 
allows us to scan a wide range of temperatures, not studied earlier, keeping the lattice volumes sufficiently 
large. This is very difficult to achieve in standard thermal lattice gauge theory simulations with fixed number 
of sites along the Euclidean time direction, with new strategies being currently investigated~\cite{Bresciani:2025mcu}. 
Our emphasis is to extract some physically relevant quantities from the QCD sphaleron rates. First, we provide 
an estimate of the thermalization time for the ultra-soft momentum modes of SU(2) and SU(3) gluons during the 
early stages of the re-heating phase by matching the sphaleron rates between a thermal and non-thermal plasma 
with equal energy densities.

By accurately calculating $\Gamma_{\text{sph}}$ we also estimate its contribution in the production rate of 
thermal (QCD) axions with momenta $\vert \mathbf{k}\vert=0, \omega\to 0$ at a wide range of temperatures.  
We also calculate for the first time, the production rate of axions with dispersion $\omega \simeq \vert 
\mathbf{k}\vert$ due to non-perturbatively interacting soft gluons which for temperatures less than the 
electroweak scale is sigificantly larger than the perturbative predictions in the hard thermal loop 
effective theory. The plan of this paper is as follows: After briefly reviewing the key aspects 
of the effective theory of magnetic gluons at high temperatures we describe our algorithm to generate 
statistically independent gauge configurations by discretizing the effective Hamiltonian on a 3D spatial 
lattice. In the subsequent section, we discuss our numerical procedure to extract the sphaleron rate 
in a SU(N) plasma in thermal equilibrium at high temperatures. Comparing these results with that of a 
non-thermal plasma of over-occupied gluons we extract typical timescales relevant for the thermalization 
of such ultra-soft gluons in the early stages of reheating epoch. We conclude by discussing two 
important physical implications of our results in the context of preheating in the early universe and 
for the relic axion yields.

\section{Lattice Implementation}
\label{sec:algorithmth}

\subsection{Algorithm for generating gauge field configurations: Thermal case}
We revisit the basic features of a finite temperature effective theory of the soft magnetic 
gluons~\cite{Bodeker:1998hm}.  In a temperature regime where the soft, semi-hard and the 
hard scales are well separated i.e. $g^2T/\pi \ll gT \ll \pi T$,  gluons with 
momenta $\gtrsim gT$ can be integrated out resulting in an effective Hamiltonian 
description~\cite{Bodeker:1998hm} of the dynamics of the soft gluons whose momenta 
are of the order of the magnetic scale and lower. The equation of motion of 
these soft gluons of a SU(N) gauge theory is denoted by
\begin{equation}
-\partial_t E_{\mathbf{x}}^{ic}+[D_j,F^{ji}(\mathbf{x})]^c=\sigma E^{ic}_{\mathbf{x}}+\zeta^{ic}_{\mathbf{x}}(t)~,
~c=1,.., N^2-1.
\label{eqn:langevinevogauge}
\end{equation}
Here $\sigma$ is the color conductivity which is known upto 
next-to-leading-log order~\cite{Arnold:1999uy},
\begin{equation}
    \sigma^{-1}=\frac{3N g^2T}{4\pi m_D^2}\left[\ln\frac{m_D}{\gamma}+3.041\right]~
    \label{eq:colorcond}
\end{equation}
in terms of the Debye mass in pure gauge theory, $m_D=\sqrt{\frac{N}{3}}gT+\mathcal{O}(g^3)$. The mean rate of color randomization $\gamma$ is estimated self-referentially 
through the relation~\cite{Arnold:1999uy},
\begin{equation}
    \gamma=\frac{Ng^2T}{4\pi}\left[\ln\frac{m_D}{\gamma}+3.041\right]~.
\end{equation}

The $\zeta^{ic}_{\mathbf{x}}(t)$ are the stochastic color-force fields with color index $c$ which 
satisfies the fluctuation-dissipation relation, 
$\langle\zeta^{ic}_{\mathbf{x_1}}(t_1)\zeta^{jd}_{\mathbf{x_2}}(t_2)\rangle=2T\sigma\delta^{ij}\delta^{cd}\delta^{3}
(\mathbf{x_1-x_2})\delta(t_1-t_2)$.
This effective Hamiltonian for the magnetic gluons describes their interactions in terms of a 
stochastic noise which mimics random kicks on them due to the hard gluons and a term proportional 
to the color conductivity which acts to dampen these large random forces. These two contrasting 
contributions from the hard modes ensure that the soft gluons attain a thermal distribution, under 
a sufficiently long enough time evolution.

Discretizing Eq.~\ref{eqn:langevinevogauge}, the electric fields, noise fields and color-conductivity 
can be written in dimensionless units  
$gE_{\mathbf{x}}^{i}a^2,~ga^2 \delta t\zeta^{i}_\mathbf{x}(t)$ and $\sigma \delta t=\frac{\sigma}{T}T \delta t$ 
respectively, where $a$ is the lattice spacing. We set $\sigma/T$ to its perturbative estimate given in 
Eq.~\ref{eq:colorcond} and numerically implement Eq.~\ref{eqn:langevinevogauge}, on a three dimensional 
lattice with a spatial size $N_s^3$ by recasting it in dimensionless units. The discretized version of 
Eq.~\ref{eqn:langevinevogauge} is solved using the leap-frog method with a time step $\delta t/a=0.01$ 
and evolved until $t \sim 2000$-$4000~\delta t$ for the algorithm to produce thermal gauge configurations.  
We next save $\sim 640$ statistically independent gauge configurations at each temperature, which are 
separated by $\delta t/a=50$ for performing thermal averages in order to extract physical quantities. Note that the two parameters that enter into Eq.~\ref{eqn:langevinevogauge}, the 
coupling $g$ and temperature $T$  are not independent but are related to each other through the 
two-loop beta function. The strong coupling $g$ is calculated at the scale $\pi T$. The Gauss law constraint 
was implemented with a precision of $10^{-15}$ at $t=0$ and it was checked to remain so at later times.

At high temperatures, occupation numbers for the gluons with momenta $\lesssim g^2T/\pi$ 
are much larger than unity, hence these interact classically. Such a classical system can be realized 
as consisting of $2(N^2-1)N_s^3$ oscillators of energy $a T$ which has an energy density on the 
lattice~\cite{kunihiro2010chaotic} given by 
$\frac{2(N^2-1)}{N_s^3}\sum_{\mathbf{k}}|\mathbf{k}|\frac{a T}{|\mathbf{k}|}~,$
where the sum is over all allowed lattice momenta $\mathbf{k}$. We have verified that the energy 
density that we compute on the lattice agrees well with this estimate. We then set the lattice spacing 
in the effective theory in physical units from the criterion that the measured energy density matches 
with its Stefan-Boltzmann value at each $T\gtrsim 600$ MeV. This results in a condition $Ta=(30/\pi^2)^{1/3}$. 
This criterion for setting the scale in the effective theory is motivated from lattice calculations of energy 
density in four dimensional SU(2)~\cite{Engels:1994xj} and SU(3) gauge theories~\cite{Boyd:1996bx, Borsanyi:2012ve} 
which quantify the deviation from ideal gas limit to be $\lesssim 18\%$ at $2~T_c$ where 
$T_c\sim 300$ MeV, to about $\sim 1\%$ at the electroweak scale. Nevertheless the magnitude of 
these deviations sets an error on our lattice spacing estimate at each temperature which ranges from $6\%$ to
$0.6\%$ at $2~T_c$ and $20~T_c$ respectively. We have checked that the sphaleron rates are insensitive to the 
variation of the lattice spacing within $1$-$\sigma$. 

Evolving the color electric fields in a thermal ensemble as a function of time, we  
perform cooling of the gauge links and electric fields at time steps $1$-$10\delta t$ 
depending on  whether the coupling is $g\gtrsim 1$ or $g<1$, in order to remove ultraviolet 
fluctuations in them. Details of the cooling procedure and the implementation of 
Chern-Simons current is explained in the subsequent sub-sections. Performing time 
evolution of color electric fields and gauge links from an initial thermal configuration, 
we measure the Chern-Simons number change as a function of the time interval of 
observation, $\Delta t=t-t_0$, where $t_0/a=100$.

\subsection{Algorithm for generating gauge field configurations: Non-thermal case}
\label{sec:algorithmNth}

In order to calculate the sphaleron rates in a non-thermal plasma, we have to first 
choose a suitable non-equilibrium initial state. We start from an initial condition 
where the gluons labeled by momentum $\mathbf{p}$ are occupied according to a phase-space 
distribution given by, 
$\tilde f(\mathbf{p})= g^2 f(\mathbf{p})=n_0\frac{Q_s}{\vert \mathbf{p} \vert}\rm{e}^{\frac{-\vert \mathbf{p} \vert^2}
{2Q_s^2}}$ where $Q_s$ is the gluon-saturation scale which is typically between 
$1$-$2$ GeV~\cite{gelis2010color} and $g$ is the gauge coupling. This is to ensure that the occupation numbers 
of soft gluons is non-perturbatively large hence their dynamics is classical. Sampling the 
gauge links and the electric fields from this initial distribution at $t=0$,  the color electric 
fields are evolved according to a Hamilton's equation which is similar to Eq.~\ref{eqn:langevinevogauge} 
but with the right hand side of it set to zero. The classical Hamiltonian evolution of the 
gauge links and color electric fields on a spatial lattice of size $N_s^3$ and spacing $a$ 
are  performed using the leap-frog integrator with a time step $\delta t=0.01a$.  At sufficiently late times, 
$Q_s.t_0\sim 50$, the scales that characterize the soft, semi-hard and hard gluons in this non-thermal 
plasma exhibit a characteristic time dependence~\cite{Berges:2023sbs}, thus separating out 
from each other. The late-time momentum distribution function of gluons exhibits a specific time 
dependence characteristic of a non-thermal fixed point~\cite{berges2014universal}, during which we 
measure the Chern-Simons number change as a function of the observation time $\Delta t=t-t_0$ by performing 
a cooling of the gauge fields at every interval $10 \delta t$, details of which is described in the next sub-section. 
The $n_0$ values that determine the initial gluon distributions are chosen such that the magnitude of the 
initial energy densities are similar to the Stefan-Boltzmann values in a thermal plasma at chosen values of 
temperature, to enable a comparison of the sphaleron rates among them. The details about the different parameters 
and the number of non-thermal configurations generated are mentioned in table~\ref{tab:table1}. We will 
henceforth denote all dimensional quantities in units of $Q_s$. 

\begin{table}[ht]
 \centering
 \begin{tabular}{|c|c|r|r|c|c|r|r|}
 \hline 
  \hline
& $Q_s.a$ & $N_s$ &  $N_{\text{confs}}$ & & $Q_s.a$ & $N_s$ &  $N_{\text{confs}}$ \\
 \hline
 SU(2)& 1.0 & 64 & 320 & SU(3) & 0.5 & 64 & 320 \\
 & 1.0 & 128 & 320  &  & 1.0 & 64 & 320 \\
  & 0.5 & 64 & 320 &  & 1.0 & 48 & 320 \\
 & 0.5 & 128 & 320 &  & 1.0 & 96 & 192 \\
  \hline
  \hline
\end{tabular}
  \caption{Parameters and statistics for non-equilibrium classical-statistical simulations of SU(2) and SU(3) gauge theories.}
  \label{tab:table1}
\end{table}

\subsection{Measuring the Chern Simons number change using cooling}

We will next outline the details of our procedure which we primarily follow from Ref.~\cite{Mace:2016svc} 
to calculate the change in the Chern-Simons number due to sphaleron transitions.  We first start with the 
gauge links and the electric fields obtained using classical Hamilton's equation at a time $t_1$, and successively 
remove the ultraviolet fluctuations of these fields through a procedure known as calibrated 
cooling~\cite{Ambjorn:1997jz,Moore:1998swa} in order to reach to the nearest vacuum configuration. 
The spatial gauge links are updated along the \emph{fictitious} cooling time direction $\tau$ according 
to the following equation
\begin{equation}
    U_i(\mathbf{x},t;\tau+d\tau)=\rm{e}^{-igaE^i_{cool}(\mathbf{x},t;\tau)d\tau}~U_i(\mathbf{x},t;\tau)~
\end{equation}
where the color components of the \emph{cooled} electric fields labeled by $c$ are defined as,
\begin{equation}
    E^{i,c}_\text{cool}(t,\mathbf{x})=-\frac{\delta H}{\delta A_i^c(\mathbf{x},t)}.
\end{equation}
In terms of the elementary plaquette variables $U_{i,j}^{\square}$ the above equation can be re-written as,
\begin{eqnarray}
    E^{i,c}_{\text{cool}}(\mathbf{x},t;\tau)&=&-\frac{2}{ga^3}\sum_{j\neq i}  \\ \nonumber
   && \text{ReTr}\Big[t^c\left[U_{i,j}^{\square}-U_{i,-j}^{\square}\right](\mathbf{x},t;\tau)\Big]~
\end{eqnarray}
where $t^c$ are the generators of SU(N). We performed successive cooling updates upto some optimal cooling time 
$\tau_c$ such that the short distance fluctuations are removed sufficiently enough to be close to a vacuum 
configuration. We then repeated the same procedure on a gauge field configuration at a different instant of 
time $t_2$, which takes it to another nearest vacuum state. In order to now calculate the difference in the 
Chern-Simons numbers between these two vacua we first need to reconstruct the connection between the 
cooled configurations $U_\mu(\mathbf{x},t_1,\tau_c)$ and $U_\mu(\mathbf{x},t_2,\tau_c)$. We implement this 
through a smooth interpolation between the electric fields defined at $t_1$ and $t_2$ and reconstructing the 
gauge links at time $t$ where $t_1\leq t \leq t_2$, according to  
\begin{eqnarray}
    \nonumber
   E_i^{t_1\rightarrow t_2}(\mathbf{x},\tau_c)&=&\frac{i}{ga(t_2-t_1)}\ln\left[U_i(\mathbf{x},t_2,\tau_c)U_i^\dagger(\mathbf{x},t_1,\tau_c)\right], \\
    U_i(\mathbf{x},t;\tau_c)&=&\rm{e}^{-igaE_{i}^{t_1\rightarrow t_2}(\mathbf{x};\tau_c)(t-t_1)}~
    U_i(\mathbf{x},t_1;\tau)~.
\end{eqnarray}
This is a well-justified procedure since the topology does not rely on the exact path 
connecting the two vacuum configurations in the gauge space. We next calculate the change in Chern-Simons 
number between these two vacuum configurations by numerically performing the time integral of the lattice 
discretized version of Eq.~\ref{eqn:CSnumberChange}, using Simpson's method and summing over all sites within 
the lattice volume, giving us
\begin{equation}
\begin{split}
    N_\text{CS}^{\tau_c}(t_2)&-N_\text{CS}^{\tau_c}(t_1) =\frac{g^2a^3(t_2-t_1)}{8\pi^2}\sum_{x,i}E_{i,\text{imp}}^{t_1\rightarrow t_2}(\mathbf{x};\tau_c) \times\\
    &  \frac{B_{i,\text{imp}}(t_1;\tau_c)+4B_{i,\text{imp}}(t_\text{mid};\tau_c)+B_{i,\text{imp}}(t_2;\tau_c)}{6}~.
\end{split}
\end{equation}
Here $t_\text{mid}=\frac{t_1+t_2}{2}$ and we use an improved definition for the Chern-Simons 
current in terms of the $\mathcal{O}(a^2)$-improved electric and magnetic fields. The improved electric fields 
are defined at each site on the lattice as
\begin{equation}
\begin{split}
   & E_{i,\text{imp}}^c(\mathbf{x}) =-\frac{1}{12}U_i^{cd}(\mathbf{x})E^d_i(\mathbf{x}+\hat{i})\\
  & +\frac{7}{12}U_i^{\dagger cd}(\mathbf{x}-\hat{i})E^d_i(\mathbf{x}-\hat{i}) \\
    \quad &+ \frac{7}{12}E_i^c(\mathbf{x})-\frac{1}{12}U_i^{\dagger cd}(\mathbf{x}-\hat{i})U_i^{\dagger de}(\mathbf{x}-2\hat{i})E^e_i(\mathbf{x}-2\hat{i})~,
\end{split}
\end{equation}
where $U^{cd}=2\text{Tr}[t^cUt^dU^{\dagger}]$. The  $\mathcal{O}(a^2)$-improved magnetic fields are constructed 
from a combination of the four elementary ($1 \times 1$) plaquettes and the eight adjacent rectangular 
($2 \times 1$) plaquettes according to,
\newcommand{\rectsym}{\fbox{\rule{0.3em}{0.1em}}}

\begin{equation}
   \begin{split}
        & B_{i,\text{imp}}^c(\mathbf{x})=\frac{\epsilon^{ijk}}{ga^2}\text{ReTr}\left[it^c \left(\frac{5}{3}\sum_{4\square}U^\square_{\pm j,\pm k}(\mathbf{x}) \right. \right. \\
        & -\left. \left. \frac{1}{3}\sum_{8\rectsym}U^{\rectsym}_{\pm j,\pm k}(\mathbf{x})\right)\right]~.
  \end{split}
\end{equation}

\section{Results}

\subsection{Sphaleron rate in a thermal non-Abelian plasma: SU(2) vs SU(3)}

\begin{figure}
    \centering
    \includegraphics[width=0.49\textwidth]{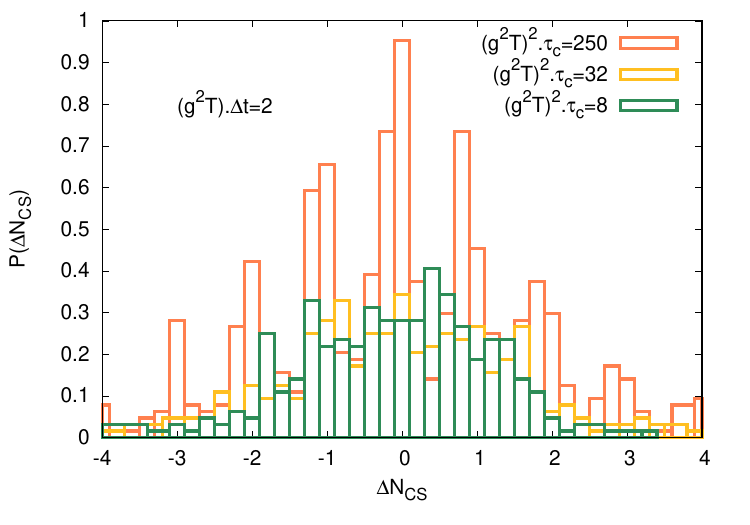}
    \includegraphics[width=0.49\textwidth]{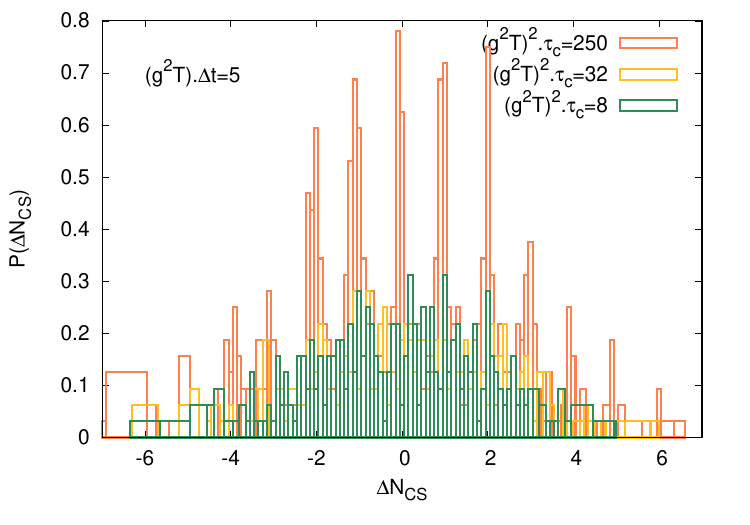}
    \caption{Probability distribution of $\Delta N_\text{CS}$ at different optimal cooling depths $\tau_c$ in 
    a thermal SU(3) gauge plasma at $g=0.58$. }
    \label{fig:prob_NCS}
\end{figure}

\begin{figure}
    \centering
    \includegraphics[width=0.48\textwidth]{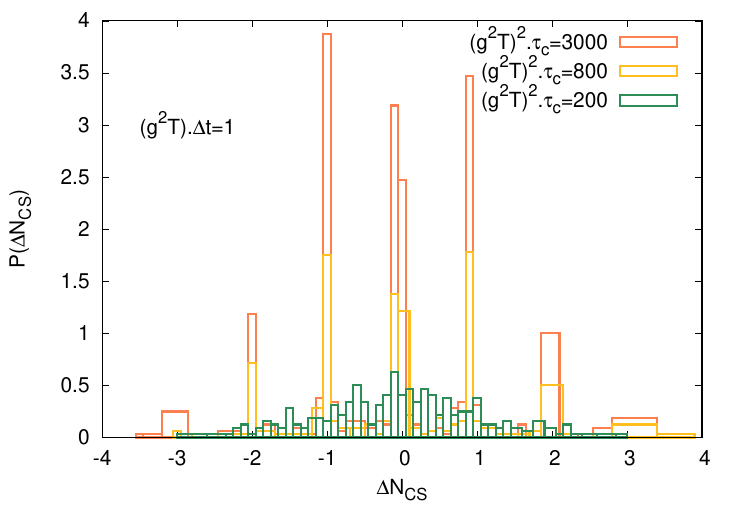}
    \includegraphics[width=0.48\textwidth]{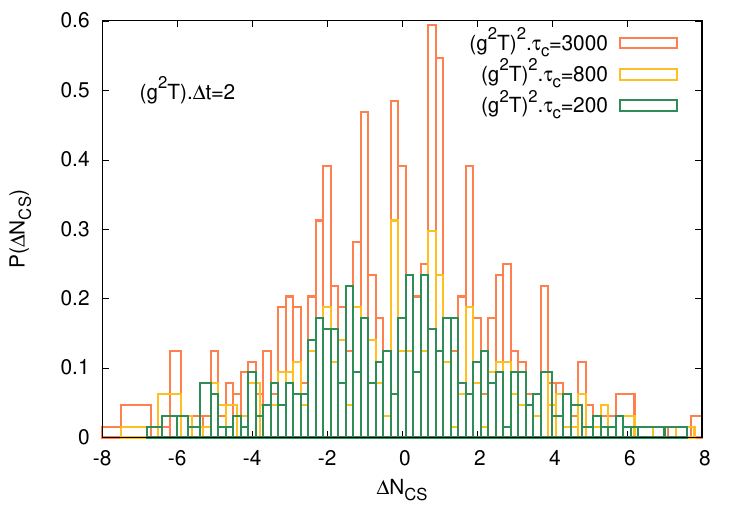}
    \caption{Probability distribution of $\Delta N_\text{CS}$ at different optimal cooling depths $\tau_c$ in 
    a thermal SU(3) gauge plasma at $g=1.12$.}
    \label{fig:prob_NCS_2}
\end{figure}

Performing the cooling procedure described in the preceeding section, we first calculate the distribution of the 
Chern-Simons number for different choices of the optimal cooling time $\tau_c$ for both SU(2) and SU(3) 
thermal plasma. In Fig.~\ref{fig:prob_NCS}, the probability distribution of Chern-Simons number change 
$\Delta N_\text{CS}$ in SU(3) gauge theory is shown for different cooling times at a very high temperature 
denoted by $g=0.58$. 
It is evident that at $(g^2T)^2\tau_c=250$, the distribution of $\Delta N_\text{CS}$ peaks close to integer values. 
The peaks are more sharply concentrated near the integer values as the system evolves in time $\Delta t$, 
starting from  $\Delta N_\text{CS}=0$ at $t=0$.  Furthermore the Chern-Simons number diffuses to larger 
values with time, which is evident from the lower panel of Fig.~\ref{fig:prob_NCS}. Under sufficiently long  
time evolution, the clustering of $\Delta N_\text{CS}$ values around integers is already visible at 
$(g^2 T)^2\tau_c\gtrsim 32$, henceforth we will perform cooling upto this optimal depth for $g<1$.  
We observe a similar trend at a lower temperature denoted by $g=1.12$ shown in Fig.~\ref{fig:prob_NCS_2}, but at a a comparatively larger cooling depth $\tau_c$, which ensures that the UV fluctuations are optimally removed and the values of $\Delta N_\text{CS}$ are peaked sufficiently close to integer values. Thus for 
gauge configurations at low temperatures denoted by $g>1$, we will usually perform cooling upto optimal 
values $(g^2 T)^2\tau_c=3$-$9\times 10^3$.

\begin{figure}
    \centering
    \includegraphics[width=0.49\textwidth]{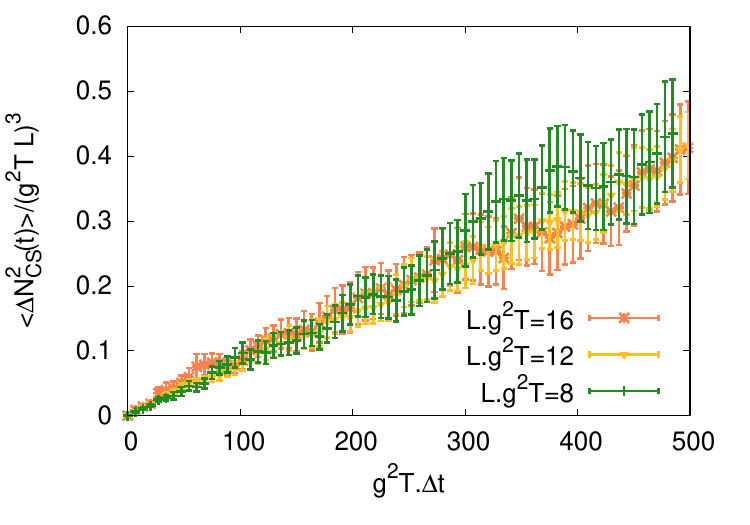}
    \caption{Autocorrelation of the Chern-Simons number change as a function of the observation time $\Delta t$ 
    in a thermal SU(3) plasma at $g=0.58$, estimated for three different lattice volumes.}
    \label{fig:Autocorr_SU3}
\end{figure}

\begin{figure}
    \centering
    \includegraphics[width=0.49\textwidth]{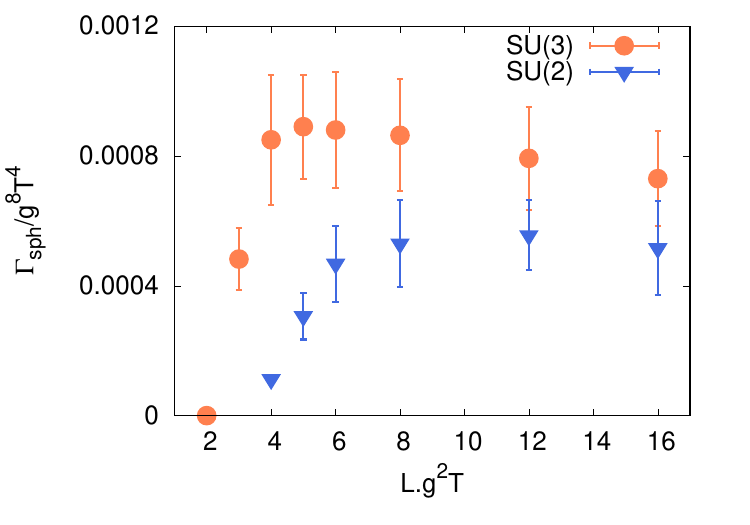}
    \caption{Volume dependence of the sphaleron rates in a thermal SU(2) and SU(3) plasma at a particular temperature 
    corresponding to $g=0.58$. The rates for SU(2) are scaled by a factor $\sim 9$ in order to show them in the same plot.}
    \label{fig:Sph_volume}
\end{figure}

We next show the autocorrelation of the Chern-Simons number change as a function of the time (interval 
of observation) $\Delta t$, in a SU(3) thermal plasma at $g=0.58$, for three different physical volumes in 
Fig.~\ref{fig:Autocorr_SU3}. It is evident that the autocorrelation of Chern-Simons number change is not 
sensitive to the physical volume when $Lg^2T\gtrsim 8$ and varies linearly with time indicating its 
diffusive nature. Extracting the sphaleron rate $\Gamma_\text{sph}$ from its slope, we next compare its 
volume dependence for SU(3) as well as an SU(2) plasma at $g=0.58$ in Fig.~\ref{fig:Sph_volume}. 
The sphaleron rate attains a plateau for spatial lengths of the lattice $Lg^2T\geq 8$, irrespective of 
the gauge group and henceforth we will calculate sphaleron rates on lattice of size $Lg^2T > 8$. We have 
also verified that the volume dependence of the sphaleron rate is under control at comparatively lower 
temperatures denoted by $g>1$ for lattice sizes $Lg^2T>8$.

\begin{figure}
    \centering
    \includegraphics[width=0.49\textwidth]{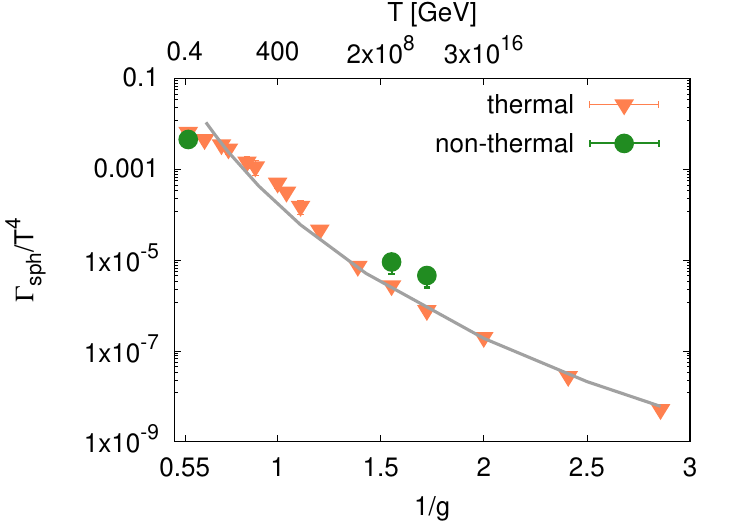}
    \caption{Sphaleron rates for a thermal SU(2) plasma as a function of increasing temperatures which is equivalently represented in terms of the inverse gauge coupling $1/g$. The solid line is the parametric dependence obtained by 
performing a fit to the lattice results of the sphaleron rate at small values of couplings in Ref.~\cite{Moore:2010jd}. Circular data points represent sphaleron rates in non-thermal plasma which has similar energy densities as the thermal case at three different temperatures.}
    \label{fig:Sph_SU(2)_eq}
\end{figure}

\begin{figure}
    \centering
    \includegraphics[width=0.49\textwidth]{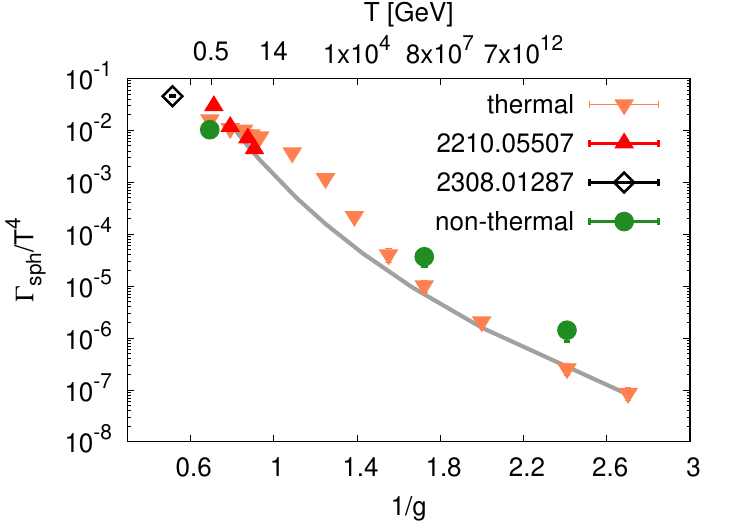}
    \caption{Sphaleron rates for a thermal SU(3) plasma as a function of increasing temperatures which is equivalently represented in terms of the inverse gauge coupling $1/g$. The solid line is the parametric dependence obtained by 
performing a fit to the lattice results of the sphaleron rate at small values of couplings in Ref.~\cite{Moore:2010jd}. Circular data points represent sphaleron rates in non-thermal plasma which has similar energy densities as the thermal case at three different temperatures. The triangles are data from Ref.~\cite{BarrosoMancha:2022mbj} in a SU(3) gauge theory and the black diamond is the data for $2+1$ flavor QCD from Ref.~\cite{Bonanno:2023thi}.}
    \label{fig:SphRate_SU3}
\end{figure}

In Figs.~\ref{fig:Sph_SU(2)_eq} and \ref{fig:SphRate_SU3} we have shown our results for the extracted sphaleron rates 
as a function of the inverse gauge coupling or equivalently temperature, which typically varies between $0.6$-$10^{15}$ 
GeV. The corresponding spatial lattice size varies between $16 \lesssim L g^2T\lesssim 108 (72)$ for $N=2(3)$ 
respectively. Our lattice results are compared with the parametric dependence of $\Gamma_\text{sph}$ obtained by 
performing a fit to the earlier lattice results of the sphaleron rate at perturbatively-weak couplings $\alpha_s=g^2/(4\pi) \ll 1$ ~\cite{Moore:2010jd}, which for SU(N) gauge group is
\begin{equation} 
{\label{pert_sph}}
    \Gamma_{\text{sph}}=0.21(1)\frac{g^2 T^2}{m_D^2}\left(\ln\frac{m_D}{\gamma}+3.041\right)\frac{N^2-1}{N}(N\alpha_s)^5T^4~.
\end{equation}
Here $\gamma$ is the damping rate which to the leading order in $g^2$ is denoted in terms of the Debye mass 
$m_D$ as $\gamma=\frac{Ng^2T}{4\pi}\left(\ln\frac{m_D}{\gamma}+3.041\right).$
Our lattice results for the sphaleron rate start to agree with the above parametric estimate at sufficiently 
weak couplings $1/g \gtrsim 1.5~ (2.0)$ for $N=2~(3)$ which corresponds to temperatures $T>10^8~(10^{10})$~GeV. 
For comparison, we also show the results for sphaleron rates in a non-thermal plasma with similar energy densities 
as the thermal case as circles in the same figure. We will provide a comparative discussion regarding these 
non-thermal data in the next section. We also perform a comparison of the existing results for the sphaleron 
rates in thermal SU(3) plasma with and without dynamical fermions from Refs.~\cite{BarrosoMancha:2022mbj} and \cite{,Bonanno:2023thi} respectively. Our results are lower in magnitude compared to results obtained in pure 
SU(3) at $T \simeq 600$ MeV~\cite{BarrosoMancha:2022mbj} but increases with temperature and eventually agrees 
with the perturbative estimates. This is due to the fact that the damping due to the hard gluons are only 
included at $\mathcal{O}(\ln g)$ in our case compared to the full theory hence our estimates for the rates 
are higher (lower) at $g>1~ (<1)$. Presence of dynamical fermions do not cause a significant enhancement of the 
sphaleron rate at a lower temperature $T \sim 350$ MeV~\cite{Bonanno:2023thi}, where the $\Gamma_\text{sph}$ is 
close to our results obtained within an effective theory. However note that our results rely on the efficient 
separation of scales in a thermal plasma, which might not be the case at this temperature.

\begin{figure}
    \centering
    \includegraphics[width=0.48\textwidth]{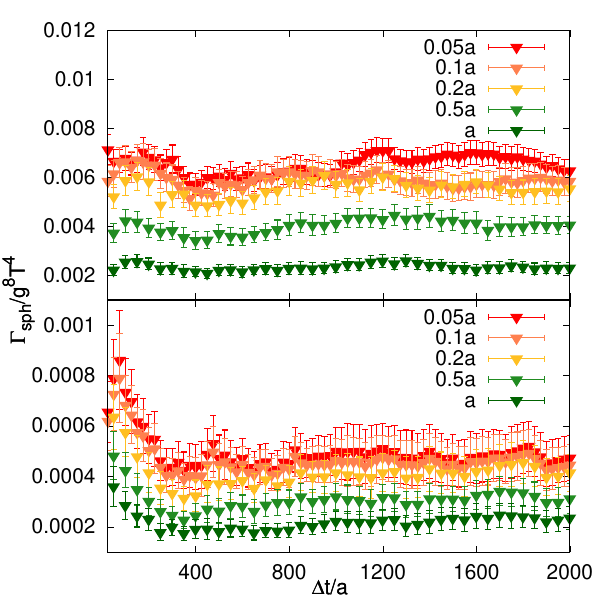}
    \caption{Sphaleron rates as a function of the  $\Delta t/a$ obtained by performing calibrated cooling of the gauge fields at time intervals ranging between $0.05$-$1.0a$ for a thermal plasma consisting of (top) SU(3) and (bottom) SU(2) gluons at $g=1$.}
    \label{fig:Sph_SU(2)_coolfreq}
\end{figure}

Before proceeding with our calculations for the sphaleron rate in the non-thermal plasma we also discuss about an 
important systematic effect arising due to the choice of interval at which the cooling of gauge fields are performed.  
In Fig.~\ref{fig:Sph_SU(2)_coolfreq}, we show our results for the sphaleron rates for an SU(2) plasma at $g=1$ as a 
function of how frequently we have performed cooling of the gauge fields during their Hamiltonian evolution. 
Evidently the sphaleron rate changes by only $1\%$ when the cooling procedure is performed at an interval 
$0.1a=10 \delta t$ as compared to $0.05a=5\delta t$. We will henceforth calculate sphaleron rates after 
performing cooling at every time step $10\delta t$, for all temperatures corresponding to $g\leq 1$. However 
at comparatively lower temperatures where $g>1$, we observe a noticeable dependence on the cooling frequency. 
Hence to minimize systematic errors at lower temperatures, we have chosen to perform cooling at each 
time-step $\delta t$ in order to extract $\Delta N_\text{CS}$. Another source of systematic error is due to finite lattice spacing artifacts. At the lowest temperature 
for which the lattice spacing $a$ is coarsest, the quantity $g^2Ta\sim 3(4)$ for $N=3(2)$ respectively. The size of the 
sphaleron on the other hand, is typically $\sim \frac{4 \pi}{g^2 T}$~\cite{McLerran:1990de}. Thus even for our coarsest 
lattice, its spacing is much smaller compared to the typical size of a sphaleron, irrespective of the number of colors. 
This ensures that discretization errors on $\Gamma_\text{sph}$ is sufficiently well-controlled for the range of 
temperatures we have studied. Comparing our estimated sphaleron rate $\Gamma_\text{sph}/T^4=0.0109\pm 0.002$ at 
$T=4~T_c$, with the continuum extrapolated rate $\Gamma_\text{sph}/T^4=0.0115\pm 0.0006$~\cite{BarrosoMancha:2022mbj} 
in 4D SU(3) gauge theory at the same temperature, we observe a perfect agreement. This reinforces that the 
discretization errors in our results are under control. 

\subsection{Extracting sphaleron rates in a non-thermal plasma: SU(2) versus SU(3)}
\label{Sec:SphRateGlasma}

\begin{figure}
    \centering
    \includegraphics[width=0.48\textwidth]{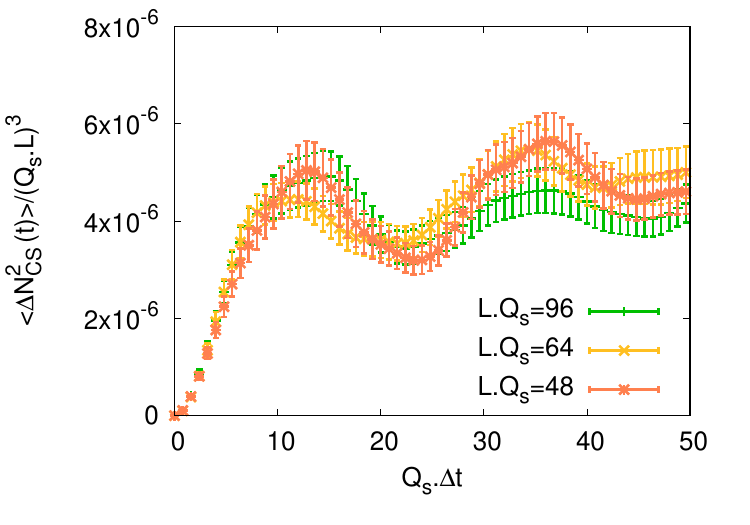}
    \caption{Autocorrelation of the Chern-Simons number change in a non-thermal SU(3) plasma for three different 
    volumes, measured as a function of the time interval $\Delta t$ for which transitions are monitored, 
    starting from an initial time $Q_s. t_0=50$.}
    \label{fig:Sph_NCs_SU3}
\end{figure}

\begin{figure}
    \centering
    \includegraphics[width=0.49\textwidth]{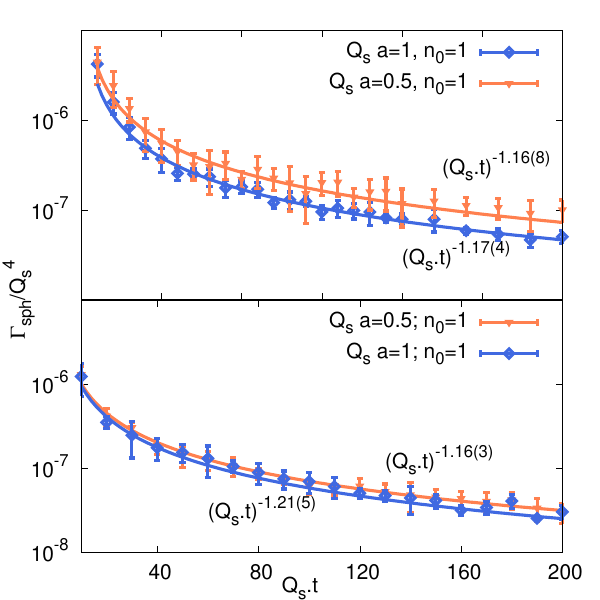}
    \caption{Parametric dependence of sphaleron rates as a function of time (at sufficiently late times) for 
    a non-thermal (top) SU(3) and (bottom) SU(2) plasma where the infrared gluons are over-occupied, shown for two different lattice with $Q_s.a=1$ (blue) and $Q_s.a=0.5$ (orange) respectively. }
    \label{fig:Sph_SU(2)_noneq}
\end{figure}

Starting from glasma-like initial conditions and evolving the gauge links using classical-statistical
algorithm described in section~\ref{sec:algorithmNth}, we calculate the auto-correlation of the Chern-Simons 
number for SU(3) gauge theory, which are shown in Fig.~\ref{fig:Sph_NCs_SU3}. The change in
$\langle N^2_\text{CS}(t)\rangle$ as a function of $Q_s.\Delta t$ is calculated for three different volumes 
$Q_sL=48, 64, 98$ and $n_0=2$ starting from an initial time $Q_s.t_0=50$. In contrast to the thermal case 
where the auto-correlation increases linearly as a function of time, we observe a linear rise in this 
case until $Q_s.\Delta t \lesssim 10$ beyond which an oscillatory pattern is observed  which is quite 
robust to different choices of lattice volumes and cooling frequencies. This observation is consistent 
with an earlier lattice study~\cite{Mace:2016svc}, where it was argued that such oscillations arise due 
to the strong collective behavior of the gluons in this over-occupied regime. We have extracted sphaleron 
rates from the slope of the initial linear growth of $\langle N^2_\text{CS}(t)\rangle$ as a function of 
$Q_s t$ for two different lattice spacings $Q_sa=1$ (blue) and $Q_sa=0.5$ (orange), which are shown in 
Fig.~\ref{fig:Sph_SU(2)_noneq} for the SU(3) and SU(2) plasma respectively. A fit to the data as a function 
of $Q_s.t$ reveals that the sphaleron rate has a parametric dependence $\sim Q_s^4 (Q_s.t)^{-1.2}$, which 
is quite robust, independent of the choice of lattice spacing or the gauge group. Incidentally the magnetic 
scale extracted from the spatial string tension also varies as $\sqrt{\sigma_s(t)}\sim Q_s(Q_s.t)^{-3/10}$ 
under sufficiently long time evolution, which ensure the onset of a non-thermal scaling 
regime~\cite{berges2014universal}.  The sphaleron rate thus parametrically behaves as 
$\Gamma_{\text{sph}}\simeq \sigma_s^2$ due to a significant contribution of gluons whose 
momenta are less than the magnetic scale. This observation is universal irrespective of the 
number of colors of the gauge group. Moreover interactions among these so-called low-momentum 
magnetic gluons are non-perturbatively large due to their large occupation numbers 
hence the sphaleron rate is intrinsically a non-perturbative quantity.

Revisiting our discussions related to the comparison of sphaleron rates between a thermal versus non-thermal plasma
shown in Figs.~\ref{fig:Sph_SU(2)_eq} and \ref{fig:SphRate_SU3}, we recall that such a comparison is done keeping the 
energy density to be the same in both cases. Within the thermal effective theory of QCD, interactions with hard gluons 
whose momenta are larger than the Debye mass scale provide damping as well as random kicks to ultra-soft gluons leading 
to an additional suppression of the sphaleron rate, otherwise determined solely by the magnetic scale, by a factor 
$\sim 1/g^2$. In the over-occupied non-thermal plasma that we study here, the hard scale increases as a 
function of time and do not influence the classical evolution of these ultra-soft modes at long enough times. 
Hence the sphaleron rates in this case are determined solely by the magnetic scale. Thus at asymptotically high 
temperatures, where couplings $g \ll 1$, thermal sphaleron rates get suppressed compared to the rates in a 
non-thermal plasma. At temperatures denoted by $g\gtrsim 1$, however the opposite trend is visible in the data, 
as expected.

\section{Physical implications of the QCD sphaleron rate }

\subsection{Estimating the thermalization time for gauge fields during early reheating era}

Inflation is one of the most attractive paradigms of early universe cosmology which can explain key 
observations~\cite{Guth:1980zm, Starobinsky:1980te,Planck:2018jri}.  At the end of inflationary epoch, the universe 
gets reheated due to the decay of the inflaton into radiation, the mechanism of which is not very well understood.  
If the coupling of the inflaton to radiation is non-perturbatively large, its decay can occur via an 
exponentially large production of soft radiation through a process called preheating~\cite{kofman1994reheating}. 
Such a non-perturbative process can also occur at weak couplings $g_\Phi$ if the amplitude of 
expectation value of the inflaton field $\Phi$ is large. In this case, the relevant parameter controlling resonant 
production of particles is $\sim \frac{g_\Phi^2 \Phi^2}{m_{\Phi}^2}$, where $m_{\Phi}$ is the mass of the inflaton. 
This ratio can be non-perturbatively large leading to efficient preheating of the universe~\cite{kofman1997towards}.
A possible scenario extensively discussed in the literature involves inflaton coupling weakly to a thermal bath of 
radiation and decaying slowly through perturbative interactions~\cite{Harigaya:2013vwa,Passaglia:2021upk,Drees:2021lbm}. 
Since the typical mass of the inflaton is large it will decay into very high energy standard model particles, 
say for e.g., gluons. Based on Boltzmann kinetic theory approach, the typical time-scale in which these high
energy gluons will lose energy to the thermal bath and acquire a thermal distribution could be significantly longer 
than the Hubble-time during the early stages of reheating, which can put strong restrictions on the maximum 
temperature achieved in the universe~\cite{Harigaya:2013vwa,Passaglia:2021upk,Mukaida:2024jiz}.

\begin{figure}
    \centering
    \includegraphics[width=0.49\textwidth]{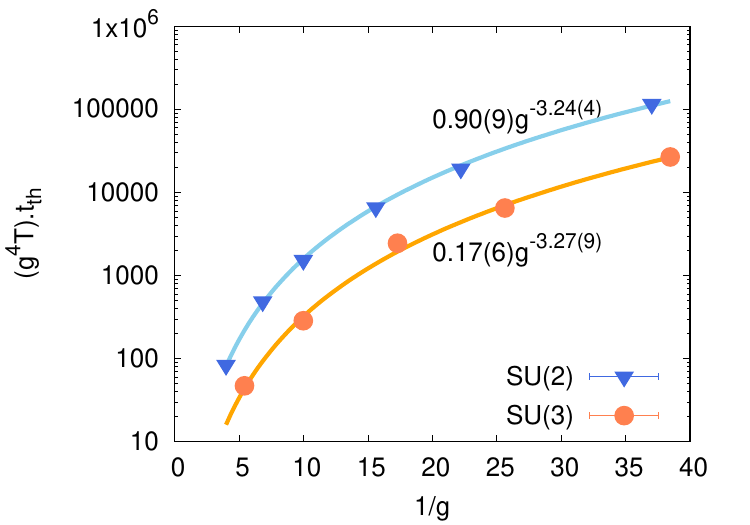}
    \caption{Parametric dependence of the time $ t_\text{th}$ required for sphaleron rates to attain their values in a thermal plasma starting from non-thermal initial conditions, conserving the energy density, shown as a function of inverse gauge coupling. }
    \label{fig:Th_Time_SU2_SU3}
\end{figure}

We would like to stress here that the conventional Boltzmann approach can only describe 
thermalization of quantum fields which allow for a quasi-particle description. The hard 
gluons in a non-Abelian plasma whose momenta are $\sim T$ can be, for e.g., described within 
kinetic theory. However it is known from classical-statistical simulations of over-occupied 
non-Abelian gauge theories that soft gluons interact non-perturbatively among themselves 
eventually thermalizing in a time-scale which is faster~\cite{Guin:2025lpy} compared to the 
typical thermalization time required for the perturbative hard modes~\cite{Baier:2000sb,Kurkela:2011ub,Fu:2021jhl}. 
It is thus important to understand the implications of any non-perturbative collective phenomena 
that might occur the very early stages of reheating~\cite{Barman:2025lvk}. We discuss one such 
possible scenario where a sphaleron dominated non-thermal SU(N) plasma comprising of non-perturbatively 
interacting over-occupied gluons, is created due to decay of the inflaton in the initial stages 
of re-heating. We then estimate the typical time required for the occupation numbers of these 
ultra-soft gluons to acquire a thermal distribution starting from a non-thermal one. As discussed 
in the previous section, the sphaleron rate in an over-occupied non-thermal SU(N) plasma can be 
typically parametrized as $\Gamma\sim Q_s^4 (Q_s t)^{-1.2}$ when the momentum distribution function 
of gluons exhibit a non-thermal scaling behavior. Comparing this with the sphaleron rate in a thermal 
plasma at a temperature $T$, one obtains a typical thermalization time $ t_\text{th}$ in units of 
$g^4T$. Results for $ t_\text{th}$ as a function of temperature or equivalently $g$, are shown in 
Fig.~\ref{fig:Th_Time_SU2_SU3}. Data points in Fig.~\ref{fig:Th_Time_SU2_SU3} can be described by 
a fit ansatz $g^{7.24(4)}T.t_{\text{th}}\simeq0.90(9)$ for SU(2) gauge theory while for SU(3) the 
analogous fit ansatz that describes our data is $g^{7.27(9)}T.t_{\text{th}}\simeq 0.17(6)$. 

The typical temperature range that signals the end of the reheating era and is compatible with the Planck data can
vary between $10^9$ and $10^{13}$ GeV~\cite{Planck:2018jri}. The gauge coupling in SU(2) typically varies between 
$g\sim0.55$-$0.65$ in this temperature range. Using now our fit function we can obtain typical time-scales 
corresponding to these couplings which denote thermalization time for the ultra-soft gluons, 
$\frac{1}{t_\text{th}}\sim 4.9\times 10^7$-$1.5 \times 10^{11}$ GeV.  For the SU(3) case, coupling typically 
varies between $g\sim0.45$-$0.53$ and thus the thermalization time of its ultra-soft modes, 
$\frac{1}{t_\text{th}}\sim5.8\times 10^7$-$1.8\times 10^{11}$ GeV. These thermalization time 
estimates are thus insensitive to the number of colors in the gauge group. Moreover we observe that $t_{\mathrm{th}} \ll H^{-1}$, where $H$ denotes the Hubble parameter governing the expansion rate of the Universe during this epoch. Hence these ultra-soft gluons undergo fast thermalization, thus providing 
a thermal bath for the hard gluons that are formed due to the perturbative decay of the massive 
inflaton~\cite{Mukaida:2024jiz}. The hard gluons further split into daughter gluons which carry 
lower momenta than their parent, within a typical time scale $t_\text{hard}$ which can be 
obtained~\cite{Mukaida:2024jiz} by matching the splitting rate $\Gamma_\text{split}$ with the 
Hubble parameter $H$, 
\begin{equation}
    t_\text{hard}\simeq \frac{(0.2-0.8)}{N^2}\Big(\frac{g^2}{4\pi}\Big)^{-16/5}\Bigg(\frac{\Gamma_\Phi}{m_\Phi^3/M_\text{Pl}^2}\Bigg)^{-3/5}m_\Phi^{-1}~.
\end{equation}
Here $\Gamma_\Phi$ and $m_\Phi$ are the decay width and mass of the inflaton respectively, whose typical estimates 
are $\Gamma_\Phi\simeq5.68\times10^{-18}M_\text{Pl}$~\cite{Barman:2025lvk,rudenok2014post,ema2016gravitational} 
and $m_\Phi=0.51\times10^{-5} M_\text{Pl}$~\cite{caravano2021lattice} in units of the Planck mass $M_\text{Pl}$. 
The mass of the inflaton is chosen to roughly match with the COBE normalization for the power spectrum of curvature 
perturbation. If the temperature during reheating is $T=10^{13}$ GeV, then $t_\text{hard}\simeq(2-8)^{-1}\times10^{-8}
~\text{GeV}^{-1}$. The time required for the ultra-soft SU(3) gluons to thermalize is less than $t_\text{hard}$ 
required by the hard gluons to entirely split into softer ones thereby acquiring a thermal distribution. Our 
findings are thus consistent with the perturbative reheating scenario. 
However if the temperatures are on the lower side $T=10^9$~GeV, then typical estimates of 
$t_\text{hard}\simeq(5.7-22.8)^{-1}\times10^{-8}~\text{GeV}^{-1}$ are such that $t_\text{hard}<t_\text{th}$, 
which is not feasible. Hence for the perturbative reheating scenario to work well which requires $t_\text{th}
<t_\text{hard}$, typical temperatures during reheating should be at least $\gtrsim 10^{10}$~GeV which is 
typically favored in Higgs inflation scenario~\cite{Bezrukov:2007ep}.

\subsection{Non-perturbative thermal axion production rate and its consequences}

Explaining the strong CP problem is one of the long-standing challenges for physics beyond the Standard Model (SM). 
Out of many proposed solutions,  Peccei-Quinn (PQ) mechanism~\cite{Peccei:1977hh,Peccei:1977ur} turns out to be one 
of the most attractive way out since it also provides a plausible explanation of the abundance of dark matter in 
the present universe. The PQ mechanism involves invoking a global $U_\text{PQ}(1)$ symmetry which is anomalous under 
the SU(3) color group of SM and is spontaneously broken at some high scale. The remnant due to this spontaneously broken 
$U_\text{PQ}(1)$ symmetry is a pseudo Nambu-Goldstone mode known as an axion, which has an anomalous coupling to 
the SU(3) color fields given by the interaction term in the Lagrangian, 
$\mathcal{L}_\text{int}=g^2/(32\pi^2 f_a)~a(x).G^a_{\mu\nu}\tilde{G}^a_{\mu\nu}(x)$. 
Here $f_a$ is the axion decay parameter, which is constrained from astrophysical and 
cosmological observations to be $4\times 10^8 < f_a < 10^{12}$ GeV~\cite{Raffelt:2006cw}. Formation 
of axion condensate provides an explanation of the cold dark matter abundance produced non-thermally through the 
misalignment mechanism~\cite{Preskill:1982cy,Abbott:1982af,Dine:1982ah}. When the primordial plasma attained a 
temperature $\sim 1 $ GeV during cosmological evolution, the axion started to feel the QCD vacuum potential 
and underwent damped oscillations around its minima as a result of which it acquired a mass whose typical values 
are $m_a=5.7~\text{eV}\times (10^6 \text{GeV}/f_a)$~\cite{GrillidiCortona:2015jxo,Petreczky:2016vrs,Borsanyi:2016ksw}. 

However there can be other mechanism of production of axions as well. Once the primordial plasma formed during 
initial stages of the evolution of our universe thermalizes, relativistic axions can be produced through scattering 
among SM particles even though they might not immediately thermalize~\cite{Graf:2010tv}. Eventually when the 
produced hot axions acquire a thermal distribution, these contribute to the radiation budget of the universe, 
quantified in terms of total number of neutrinos. The effective number of neutrinos $N_\text{eff}$ quantifies 
how many massless relativistic matter degrees of freedom contribute to the total radiation density. Any 
relativistic particle with a substantial energy density e.g., axions will contribute to $N_\text{eff}$. 
Deviation from the standard cosmological value for the number of neutrino species 
$N_\nu \simeq 3.044$~\cite{akita2020precision,froustey2020neutrino,bennett2020towards} can be estimated to be,
\begin{equation}
    \Delta N_\text{eff}=N_\text{eff}-N_\nu=\frac{8}{7}\left(\frac{11}{4}\right)^{4/3}\left(\frac{\rho_a}{\rho_\gamma}\right)_{\text{CMB}}.
    \label{eqn:DefNeff}
\end{equation}
Here $\rho_a$ and $\rho_{\gamma}$ are the energy densities of the axions and photons respectively during the 
recombination era. Any additional source of radiation should be thus detectable in the cosmic microwave background 
(CMB). Furthermore, under the assumption of instantaneous decoupling of the thermal population of axions, the above 
equation can be written as~\cite{Bouzoud:2024bom},
\begin{equation}
    \Delta N_\text{eff}\simeq\frac{4}{7}\Big(\frac{11}{4}\Big)^{4/3}\left[\frac{g_{*,s}(T_\text{CMB})}{g_{*,s}(T_D)}\right]^{4/3}~.
    \label{eq:DNeff}
\end{equation}
The decoupling temperature $T_D$ defines the epoch at which the thermal axion production rate $R(T_D)$ 
matches with the Hubble expansion parameter. Thus in order to estimate the decoupling temperature and 
$\Delta N_\text{eff}$ very precisely, one needs to accurately calculate the thermal production rate of axions. 
We consider the coupling of the axions to SU(N) gluons, where we have studied $N=2, 3$ respectively. It is well 
known that even at asymptotically high temperatures, the soft gluons in non-Abelian gauge theories whose 
momenta are at and below the magnetic scale $g^2T/\pi$, interact non-perturbatively~\cite{Linde:1980ts}. 
We thus calculate the axion production rate on lattice to accurately take into account the contribution of 
magnetic gluons, within the effective theory.

\begin{figure}
    \centering
    \includegraphics[width=0.49\textwidth]{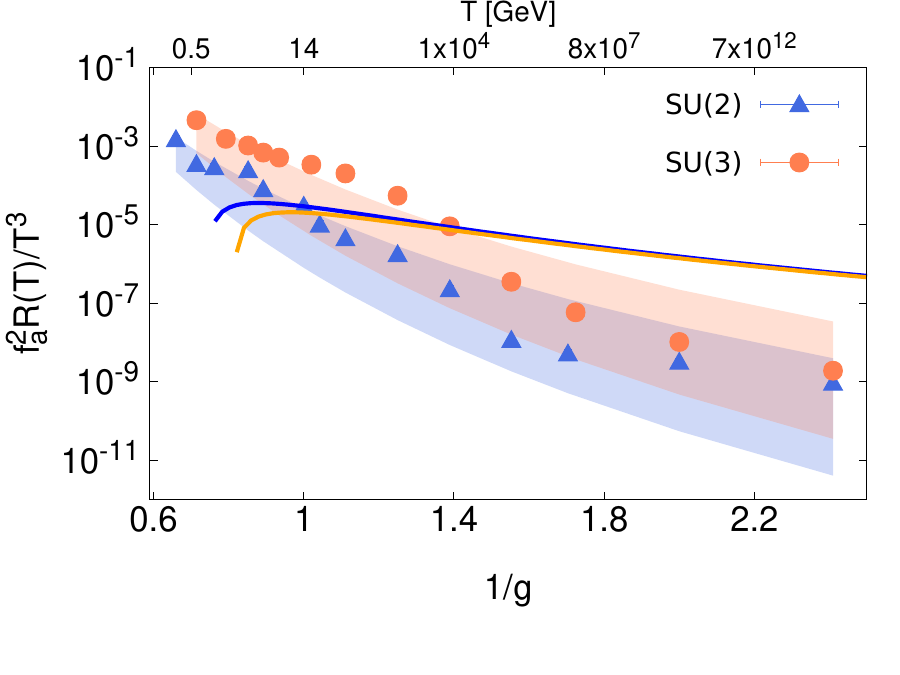}
    \caption{The thermal axion production rate in non-Abelian gauge theory shown as a function of the inverse gauge coupling. The solid lines correspond to the rates calculated perturbatively within the hard-thermal loop (HTL) resummation scheme. The shaded bands correspond to the axion production rates only due to sphaleron transitions.}
    \label{fig:SU2_axion}
\end{figure}

In thermal equilibrium, the creation and annihilation rates for axions at a temperature $T$ denoted as 
$\Gamma^<(T)$ and $\Gamma^>(T)$ respectively, are related through,
\begin{equation}\label{axionrate}
    \Gamma^>(\mathbf{k},T)=e^{\omega/T}\Gamma^<(\mathbf{k},T)=\frac{\Gamma^>_\text{top}(\mathbf{k},T)}{2\omega f_a^2}
\end{equation}
where  $\Gamma^>_\text{top}$ is extracted from the two-point correlation function of the topological 
charge,
\begin{equation}
    \Gamma^>_\text{top}(\mathbf{k},T)=\left(\frac{g^2}{32\pi^2}\right)^2\int d^4x~\rm{e}^{ik^\mu x_\mu}\Big\langle G\Tilde{G}(x)\cdot G\Tilde{G}(0)\Big\rangle~.
\end{equation}
Here $G\Tilde{G}(x)=G^a_{\mu\nu}(x)\tilde{G}^a_{\mu\nu}(x)$ and the energy of the axion $\omega$ or equivalently 
$k_0=\sqrt{\mathbf{k}^2+m_a^2}$. We consider the on-shell axions and calculate their 
production rate. Although we are using an effective theory of soft gluons where the typical momenta and frequencies are 
$\vert\mathbf{k}\vert\lesssim g^2T/\pi, ~\omega\lesssim g^4T$ respectively, the dominant contribution to the axion 
production rate due to gluons at temperatures lower than the electroweak scale, indeed comes from the soft magnetic 
sector~\cite{bouzoud2026energy}. We have calculated $\Gamma^>_\text{top}(\mathbf{k},T)$ for a wide range of 
temperatures for both SU(2) and SU(3) gauge theories on a $N_s=32$ lattice, performing a thermal average over 
$200$ configurations. Next we calculate the average thermal production rate of axions defined as,
\begin{equation}
    R(T)=\frac{1}{n_a^\text{eq}}\int\frac{d^3\mathbf{k}}{(2\pi)^3}~\Gamma^<(\mathbf{k},T)~
    \label{eqn:axionrate}
\end{equation}
by performing the momentum integration of $\Gamma^<(\mathbf{k},T)$ numerically. The lower and upper limits of the 
integral are given by $\vert\mathbf{k}\vert/T = 0, \sqrt{3}\pi/1.45$ respectively, the later is the maximum allowed 
magnitude of momentum for the particular lattice spacing that we have considered. The integration was performed 
numerically using Simpson's rule.
The quantity $n^\text{eq}_a=\zeta(3)T^3/\pi^2$ is the number density of a non-interacting gas of axions at a 
temperature $T$. Results for the rate $R(T)$ obtained from our lattice computations are compiled in 
Fig.~\ref{fig:SU2_axion}. 

Nevertheless, our results reinforces a previous study of the momentum and frequency dependence of 
$\Gamma^{>}_\text{top}$, where it was observed that the maximum contribution to the production rate 
of light-like axions characterized by $\vert \mathbf{k}\vert\sim\omega$ comes from the soft magnetic 
sector~\cite{bouzoud2026energy}.

We also compare our results with the perturbative estimates 
$R(T)=\frac{\zeta(3)g^6T^3}{64\pi^7f_a^2} \left[\log\frac{3T^2}{m_D^2}+0.406\right]$, 
where $m_D=g T\sqrt{\frac{N+n_f/2}{3}},~n_f=6, N=2$-$3$, calculated~\cite{Graf:2010tv} 
within hard thermal loop (HTL) resummation scheme, shown as solid lines in the same 
figure. At very high temperatures i.e. $g<0.66$, the perturbative HTL result is 
significantly higher than the axion production rate due to the magnetic gluons. 
But at comparatively lower temperatures, the dominant contribution to the production 
rate comes from the soft magnetic sector. Even at the electroweak scale the magnetic gluons 
contribute to almost $50\%$ of the axion production rate justifying the necessity of a non-perturbative 
calculation. The bands in Fig.~\ref{fig:SU2_axion} denote the axion production rate calculated by 
substituting $\Gamma_\text{top}^>=\Gamma_\text{sph}$ in Eq.~\ref{eqn:axionrate} and integrating over 
all momenta lower than a maximum cutoff $\vert \mathbf{k}\vert<\vert \mathbf{k}_s \vert$. The lower 
boundary of the shaded band is obtained for a typical choice $|\mathbf{k}_s|/T=3g^2/4\pi$ whereas the 
upper boundary is obtained when the cut-off $|\mathbf{k}_s|/T=\sqrt{3}\pi/1.45$. The non-perturbative 
axion production rate only due to sphalerons explains our lattice data quite well for $g<0.7$, above 
which the agreement is barely visible at upper edge of the error band. This also provides a 
compelling motivation for our calculations.

Our computation of the axion rate does not capture the contributions due to semi-hard and hard gluons. We thus add the 
hard thermal loop contribution to our results for the rate arising due to the soft magnetic gluons for $g \lesssim 1$. 
For $g \gtrsim 1$, we consider only the production rate obtained from our lattice calculation as the HTL approximation 
breaks down. We can now estimate the decoupling temperature $T_D$ at which the produced axions are no longer in thermal 
equilibrium with the bath consisting of soft SU(3) gluons by matching $R(T_D)=H(T_D)$, which comes out to 
$T_D=8\times10^4$~GeV for $f_a=4\times10^8$~GeV. At this temperature, the perturbative HTL rate is dominantly larger 
compared to the contribution only from the soft sector. As the universe evolves towards lower temperatures, the non-
perturbative production rates start to dominate and the produced axions again attain a thermal distribution.

Having obtained the thermal axion rate, we can now study the time evolution of the axion yields 
$Y(x)=n_a(x)/s(x)$, where $n_a(x), s(x)$ are the number density of axions and total entropy density 
respectively at a scaled temperature $x=T_f/T$ by solving the Boltzmann equation 
\begin{equation}
    \frac{dY(x)}{d\log~x}=\Big[Y^\text{eq}-Y(x)\Big]\frac{R (x)}{H}\Bigg[1-\frac{1}{3}\frac{d\log~g_*}{d\log~x}\Bigg]
  \label{eq:Boltzmann}  
\end{equation}
in the universe expanding with a Hubble rate $H$. Here $T_f$ is the final temperature upto which 
this evolution is performed. We have considered the fact that the number of relativistic degrees 
of freedom $g_*$ also evolve as a function of $x$. Note that the yield of thermal relic axion in the 
present universe is $Y_\text{eq}=n_a^\text{eq}/s\simeq2.6\times10^{-3}$~\cite{Graf:2010tv}. This 
evolution equation of the axion yields is obtained~\cite{masso2002axion,bernstein1985cosmological,turner1987early} 
by integrating the kinetic Boltzmann equation over the allowed momenta whose interaction kernel has 
contributions from both production and decay rates, under the assumption that phase-space distribution 
of axions follow $f({\mathbf{k}})=\frac{n_a}{n^{\text{eq}}_a}f^{\text{eq}}({\mathbf{k}})$~\cite{notari2023improved}. 
We start this non-equilibrium evolution at $T_D$ and evolve down to $T_f=600$~MeV upto which our 
effective theory calculations of $R(T)$ remain valid. The axion energy density at this final scaled 
temperature $x_f$ is $\rho_a(x=x_f)=\frac{\pi^{14/3}}{30}\left(\frac{sY(x_f)}{\zeta_3}\right)^{4/3}$
where $s=\frac{2\pi^2g_*T^3}{45}$. Now substituting 
$(\rho_a)_\text{CMB}=\rho_a(x_f)\Big(\frac{g_{*,\text{CMB}}}{g_{*,f}}\Big)^{4/3}$ in Eq.~\ref{eqn:DefNeff}
one obtains $\Delta N_\text{eff}=0.034$ for $f_a=4\times10^8$~GeV. Here the effective degrees of freedom are 
$g_{*,\text{CMB}}=3.93$~\cite{escudero2025fast} at the CMB epoch and $g_{*,f}=61.75$ 
at $T_f=600$~MeV~\cite{Borsanyi:2016ksw}. The value of $\Delta N_\text{eff}$ that we obtain is well below the 
bound $\Delta N_\text{eff}\leq0.28$ from PLANCK~\cite{aghanim2020planck}. Note that our estimate of 
$\Delta N_\text{eff}$ is slightly lower than the result reported in~\cite{bouzoud2026energy}. Our determination of the 
axion production rate is not available at lower temperatures, where a clear separation of scales no longer exists and 
thus the effective theory is not valid. As the temperature decreases further, strong interactions become increasingly 
important, leading to an enhancement of $\Delta N_\text{eff}$. Thus an accurate determination of the axion production 
rate for on-shell axions near the QCD crossover will be important in future to tightly constrain the value 
of $\Delta N_\text{eff}$.

\begin{figure}
    \centering
    \includegraphics[width=0.48\textwidth]{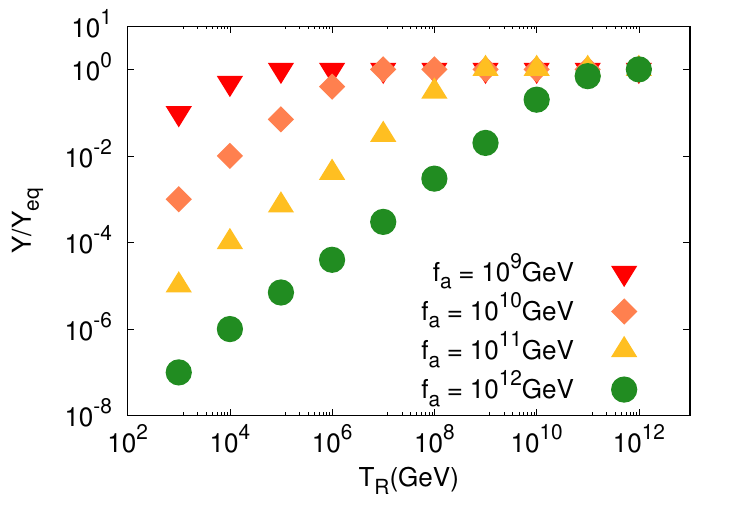}
    \caption{Present-day relic axion yields normalized by its thermal value in a SU(3) plasma obtained after solving 
    Eq.~\ref{eq:Boltzmann}, shown as a function of different initial choices of reheating temperatures 
    $T_R$  and for different allowed values of the axion decay parameter $f_a$.}
    \label{fig:axion_density}
\end{figure}

Ratios of the relic axion yields with respect to their values in thermal equilibrium in the present universe 
can be obtained by the solving Eq.~\ref{eq:Boltzmann}, starting from different initial temperatures, $T_R$ 
and allowed values of $f_a$, results of which are compiled in Fig.~\ref{fig:axion_density}. We can now 
estimate the typical decoupling temperatures $T_D$ from the onset of a kink in $Y/Y_\text{eq}$ in 
Fig.~\ref{fig:axion_density} which signals departure of the axion yields from their equilibrium values. The 
decoupling temperatures are observed to vary between $T_D\sim 10^5$- $10^{12}$ GeV depending on the values 
of $f_a \sim 10^9$-$10^{12}$ GeV. The axions thus never attain thermal equilibrium if initial reheating temperatures 
are $T_R<10^{5}$ GeV for any allowed values of $f_a$.  From our previous discussion it turns out that the correct 
hierarchy between thermalization times of the hard and soft gluons is ensured if $T_R> 10^{10}$ GeV. Given this 
scenario the thermally produced axions will fall out of equilibrium much faster for $f_a$ in the higher range 
of its of the allowed values compared to a lower range.

\section{Summary \& Outlook}

In this work, we have calculated the sphaleron transition rate in SU(2) and SU(3) gauge theory under both in and 
out-of-thermal equilibrium conditions using lattice gauge theory techniques. At very high temperatures where the 
hard, electric and magnetic scales are well separated, we have used an effective theory for the soft gluons to 
calculate the sphaleron rates for a wide range of temperatures spanning from $0.6$-$10^{15}$ GeV, without being 
beset by finite volume corrections. The sphaleron rates start to agree with their parametric estimates at weak 
couplings only at temperatures beyond the electroweak scale. Incidentally sphaleron rates in a non-thermal plasma 
are higher than in a thermal plasma with similar energy densities, where the non-thermal plasma consists of gluons 
whose occupation numbers exhibit a self-similar scaling. At high temperatures, where the hard and the soft scales are sufficiently wide apart, one would expect little effect on the sphaleron rate, which is driven primarily by soft gluons, due to dynamical quarks whose momenta are hard. In the context of the thermal effective theory we are studying, effects due to dynamical quarks enter through color conductivity, which dampens the random motion of soft gluons. Clearly the contribution due to quarks in the sphaleron rate will become significant as one approaches closer to the crossover temperature in QCD where the separation of scales is no longer valid.

By comparing the sphaleron rates in a thermal versus a non-thermal plasma at similar energy densities, we have estimated 
typical thermalization times for these ultra-soft magnetic modes during the early stages of re-heating epoch. For 
temperatures $T\geq 10^{10}$~GeV, our results demonstrate that the thermalization time for ultra-soft magnetic 
gluons is much lower compared to the timescale required for the splitting of hard gluons into softer 
counterparts obtained within the kinetic theory framework. Our results thus support a scenario where the soft 
gluons formed due to the decay of inflaton thermalizes very quickly as a result of non-perturbative interactions 
which then forms a thermal bath for hard gluons to interact with and eventually thermalize on a longer time-scale. 
Our study also allows us to put a lower bound on the reheating temperature $T_R\geq 10^{10}$~GeV for this perturbative 
reheating scenario to be a valid physical description of the early universe.

We have also calculated the axion production rate at high temperatures in SU(2) and SU(3) thermal plasma. We 
observe that the soft magnetic sector contributes significantly to the axion production rate, even at the 
electroweak scale, where our lattice results exceed the perturbative HTL prediction by a substantial margin. 
This is a generic feature of all physical observables which have a notable contribution from the non-perturbatively 
interacting magnetic gluons. We could also uncover the reason behind the large enhancement of our estimates for the 
axion production rate compared to HTL predictions at couplings $g\gtrsim 2/3$, arising due to a sizeable contribution 
from sphaleron transitions. Solving Boltzmann equation with the non-perturbative axion production rates as an input we 
could calculate the yields for relic axions.  For the allowed range of values of $f_a$, the axions typically remain in 
thermal equilibrium with the bath consisting of soft gluons for a longer time if the values of $f_a$ are on the lower 
side.  Moreover for values of $f_a$ even in the higher side of its allowed range, the decoupling temperature of 
axions is consistent with our estimated lower bound on the reheating temperature. We would in future like to 
develop new lattice techniques to calculate the axion production rate at lower temperatures, closer to the QCD 
crossover.

\section*{Acknowledgements}
We are grateful to Dietrich B\"{o}deker for a careful reading of the first draft of this work 
and for the insightful suggestions and comments. We also thank Jacopo Ghiglieri for bringing to our attention their work~\cite{bouzoud2026energy} and for the helpful suggestions. We acknowledge support
from the Institute of Mathematical Sciences, and the computing time allocation at the Institute cluster.

\bibliography{Paper_SG}

\end{document}